# A Quantum Mechanical Description of Photosensitization in Photodynamic Therapy using a Two-Electron Molecule Approximation

Vincent M. Rossi

Washburn University Department of Physics & Astronomy, Topeka, KS 66621

vincent.rossi@washburn.edu

ABSTRACT

A fundamental, Quantum Mechanical description of photoactivation of a generic photosensitizer and the ensuing transfer of energy to endogenous oxygen as part of the Type II pathway to photodamage during photodynamic therapy (PDT) is presented. The PS and molecular oxygen are approximated as two-electron molecules. Conservation of energy and of angular momenta of the two molecule system are abided via selection rules throughout the four-stage process, including initial states, absorption of a photon by the PS, conversion of the PS to an excited spin triplet via intersystem crossing (ISC), and the transition of molecular oxygen to an excited spin singlet state via a Triplet-Triplet Exchange of electrons with the PS. The provided description of photosensitization will provide students and researchers with a fundamental introduction to PDT, while offering the broader population of Quantum Mechanics and Physical Chemistry students an advanced example of quantum systems in an applied, medical context.

*Keywords: Photosensitization, Photodynamic Therapy (PDT), photochemistry, Dexter Exchange, Triplet-Triplet Exchange*

INTRODUCTION

Photodynamic therapy (PDT) is a localized and selective therapy that operates on principles included under the generic classifications of photobiology, photochemistry and photophysics (Jacques 1992; Henderson and Dougherty 1992; Hamblin and Mroz 2008; Bonnett 2000; Hasan, Moore and Ortel 2000). While PDT has found its broadest application and research as a cancer therapy, it has also been used for antimicrobial therapy for combating antibiotic resistant strains (Wainwright 1998). Three ingredients are required for PDT—a photosensitizer (PS), light, and oxygen—in order to induce photochemical damage to its targets. In short, the PS is administered to the patient and after an appropriate time interval, the targeted site is illuminated with light of appropriate wavelength to be absorbed by the PS. Upon excitation by light of appropriate energy, the excited PS interacts with endogenous molecular oxygen in order to create reactive oxygen species (ROS). The interactions between the excited PS and endogenous molecular oxygen to generate ROS has been recognized and developed over some time (Kautsky 1939; Keszthelyi et al. 1999). These ROS then interact with their immediate environment, creating oxidative damage. Targeted cancer cells or bacteria are eliminated once they reach a threshold of damage via ROS (Nilsson, Merkel, and Kearns 1972; Schmidt and Bodesheim 1998).

Absorbed photons transfer discrete energies to the PS, raising it from the singlet ground state ($^1PS$) to an excited singlet state ($^1PS^*$),

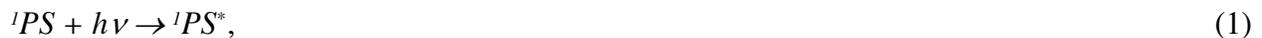

$$^1PS + h\nu \rightarrow {}^1PS^*, \qquad\qquad\qquad\qquad\qquad\qquad\qquad\qquad (1)$$

where the product of Planck's constant ($h$) and the frequency of light absorbed ($\nu$) represents the addition of energy via absorption (Fig. 1). The PS may then fluoresce back to its ground state.

Preferably, the PS in its excited singlet state will transition to its excited triplet state ($^3PS^*$) through Intersystem Crossing (ISC),

$$^1PS^* \rightarrow {^3PS^*}. \tag{2}$$

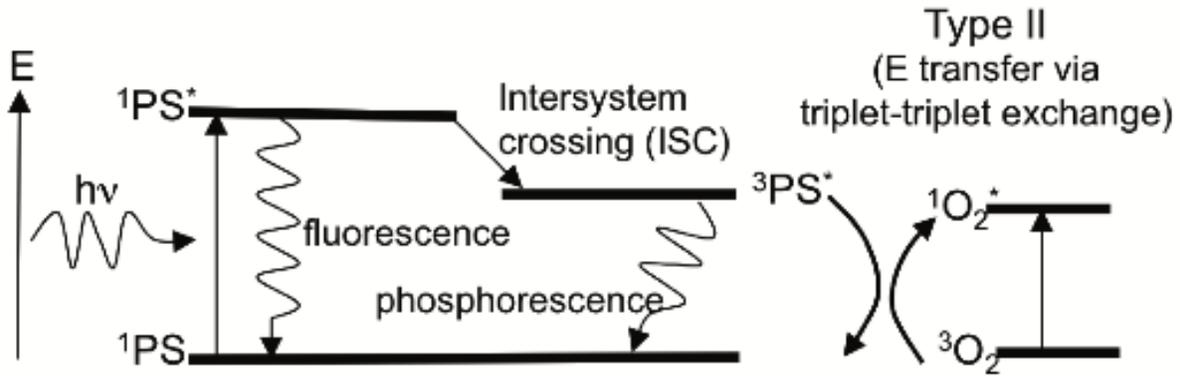

Figure 1. The process leading to the preferred Type II path to photodamage starts when the PS is excited by incident light of energy $h\nu$. The PS then relaxes via ISC to an excited triplet state, whereby it can transfer energy to molecular oxygen via a triplet-triplet electron transfer.

Once in the excited triplet state, the photosensitizer may then decay back to its ground state through one of two mechanisms. The first of which, called the Type I pathway to photodamage in PDT, involves the PS in its excited triplet state interacting with the surroundings, thereby losing energy and creating free radicals. The resulting free radicals may then react with endogenous oxygen to form cytotoxic species such as $OH^-$ (Jacques 1992; Wainwright 1998; Ochsner 1997; Peavy 2002; Prasad 2003; Mata et al. 2006).

The Type II pathway to photodamage in PDT entails a direct interaction between the PS in its excited triplet state and endogenous molecular oxygen in its triplet ground state ($^3O_2$). Such interactions, termed a Triplet-Triplet Exchange, can also cause the PS agent to decay back to its singlet ground state, in turn raising the molecular oxygen to an excited singlet state ($^1O_2^*$),

$$^3PS^* + {}^3O_2 \rightarrow {}^1PS + {}^1O_2^*. \tag{3}$$

The excited singlet state of molecular oxygen can then cause damage to its surroundings (Nilsson, Merkel, and Kearns 1972; Schmidt and Bodesheim1998). Due to the long lifetime of the excited triplet PS, sufficient time is allowed for interactions with endogenous oxygen. For this reason, the Type II pathway is generally accepted as the most common pathway to photodamage in PDT (Jacques 1992; Henderson and Dougherty 1992; Hamblin and Mroz 2008; Wainwright 1998; Ochsner 1997; Peavy 2002; Prasad 2003; Mata et al. 2006).

The above introduction to PDT is given in a typical fashion as would be found in biological or medical descriptions of PDT (Kearns and Khan 1969). The remainder of this paper is interested in giving a more rigorous, quantum mechanical explanation of the process of photosensitization in PDT. The Quantum Mechanical processes involved in activation of the Type II pathway to photodamage will be covered in a simplified fashion so as to serve as an accessible description to students and researchers who are new to PDT research. The subset of researchers responsible for light delivery and light-tissue interactions in PDT may find this description useful. As such, the quantum notation more familiar to physicists will be used moving forward. In particular, quantum states of the PS and molecular oxygen will be treated as those of two-electron molecules. Representation of photosensitization in PDT using this notation will be more familiar to the students of quantum mechanics and physical chemistry while simultaneously appealing to a rigorous sensibility by detailing the physical phenomena associated with each step of the photosensitization process (Sec. 2). The addition of angular momentum between the two molecules will be employed in order to define the overall state of the system of molecules at each step. The larger discussion will be summarize at the end of the paper (Sec. 3).

QUANTUM TWO-ELECTRON MODEL

We will consider a basic quantum mechanical example of a generic PS interacting with molecular oxygen as part of the desired Type II pathway to photodamage achieved in PDT. A generic diagram of the photocativation of the PS and interactions with molecular oxygen are depicted in Fig. 1. In particular, this work is concerned with describing the interactions between the PS and molecular oxygen from the time the PS is excited via absorption of a photon through the transfer of energy to molecular oxygen via a Triplet-Triplet Exchange. As such, all other pathways will be ignored.

Both the PS and molecular oxygen can be approximated as two-electron molecules. For example, molecular oxygen forms via the covalent bond between two oxygen atoms, each needing a pair of *2p* electrons in order to fill the *2p* shell (Turrens 2003). This pair of shared *2p* electrons will therefore be considered as those undergoing the transitions that follow during the PDT process. The same assumption will be made of the PS, considering that the exchange of energy between the PS and molecular oxygen comes in the form of electron exchange between a pair of two electron systems.

In quantum mechanics, we are concerned with eigenvalue problems where we can determine the given set of eigenstates corresponding to a given set of eigenvalues. The eigenstate of a system corresponds to the wavefunction of the system, or generically speaking, the state of the system.

The eigenvalue corresponds to some physically measureable quantity, or characteristic of the system, such as its energy, spin or angular momentum. As alluded to here, the characteristics of a quantum state can have spatial and spin dependencies, such that their corresponding wavefunctions must also incorporate spatial and spin states. We can separate the overall wavefunction, $\Psi(\vec{r}, m_s)$, into the product of the two functional dependencies,

$$\Psi(\vec{r}, m_s) = \Phi(\vec{r})\chi(m_s), \tag{4}$$

where $\vec{r}$ represents the three dimensional spatial dependence of the spatial wave function $\Phi(\vec{r})$ and $m_s$ is the spin quantum number, representing the spin dependence of the spin wavefunction $\chi(m_s)$. In this context of atomic and molecular physics, the wavefunction represents the overall state of an electron. Since electrons are Fermions, their overall wavefunctions must be antisymmetric.

When looking specifically at the context of PDT, we are dealing with systems of two electron molecules. Therefore, the overall wavefunction (4) for both the PS and molecular oxygen must be modified to reflect a two electron system,

$$\Psi(\vec{r_1}, m_{s1}; \vec{r_2}, m_{s2}) = \Phi(\vec{r_1}, \vec{r_2})\chi(m_{s1}, m_{s2}), \tag{5}$$

where the subscripts 1 and 2 represent the two separate electrons.

From the requirement for electrons to have antisymmetric wavefunctions follows the definition of the singlet and triplet states, which refer specifically to the spin wavefunction, $\chi(m_{s1}, m_{s2})$, of the two electron system. A combination of these two electrons in the spin state lead to a set of three possible symmetric wavefunctions,

$$\chi(m_{s1}, m_{s2}) = \begin{cases} \chi_{++} \\ \frac{1}{\sqrt{2}}(\chi_{+-} + \chi_{-+}) \\ \chi_{--} \end{cases} \quad (6)$$

where the + and - refer to the different combinations of spin up ($m_s = +\frac{1}{2}$) and spin down ($m_s = -\frac{1}{2}$) states, respectively. This state is specifically called the (spin) triplet state because there is a set three possible symmetric combinations for the two electron system. Similarly, there is only a single antisymmetric combination of spins,

$$\chi(m_{s1}, m_{s2}) = \frac{1}{\sqrt{2}}(\chi_{\pm} - \chi_{\mp}), \quad (7)$$

which is therefore referred to as the (spin) singlet state (Sakurai 1994).

One of the spin states from (6) or (7) can therefore be applied directly within the overall two electron wavefunction (5) for either the PS or molecular oxygen. This leaves us to more thoroughly define the spatial state of the system (Sakurai 1994). Resolving the spatial wavefunction will be based upon the quantum mechanical rules for dealing with systems of identical particles and the assumption that we can start from the model of the most simple of two electron systems---the helium atom. Under this premise, the spatial wave function can undergo a swap of electrons such that,

$$\Phi(\vec{r}_1, \vec{r}_2) = \frac{1}{\sqrt{2}}\left[\psi_{100}(\vec{r}_1)\psi_{nlm}(\vec{r}_2) \pm \psi_{100}(\vec{r}_2)\psi_{nlm}(\vec{r}_1)\right], \quad (8)$$

where the wavefunctions $\psi_{100}$ and $\psi_{nlm}$ refer to electrons in the ground and possible excited states, respectively. The two states $\psi_{100}(\vec{r}_1)\psi_{nlm}(\vec{r}_2)$ and $\psi_{100}(\vec{r}_2)\psi_{nlm}(\vec{r}_1)$ account for a change of state via exchange of identical particles—changing the configuration of the system by exchanging the states of two electrons translates to a change of state. However, the total spatial state (8) is the superposition of these two states, which can be gained either by the addition or subtraction of the two combinations. The addition of these two spatial states results in a symmetric spatial wave function. Conversely, the subtraction of the two states results in an antisymmetric spatial wave function.

Now that the symmetric and antisymmetric representations of the spatial and spin states are defined, we look to their possible combinations for the overall wavefunction of the two electron system (Sakurai 1994). Since the electron wavefunction must have overall antisymmetry, the antisymmetric spin singlet state (7) must pair with the symmetric spatial state (8), giving the overall antisymmetric singlet state $\Psi_{singlet}(\vec{r_1}, m_{s1}; \vec{r_2}, m_{s2})$,

$$\Psi_{singlet} = \frac{1}{\sqrt{2}}\left[\psi_{100}(\vec{r}_1)\psi_{nlm}(\vec{r}_2) + \psi_{100}(\vec{r}_2)\psi_{nlm}(\vec{r}_1)\right] \times \frac{1}{\sqrt{2}}(\chi_{+-} - \chi_{-+}). \quad (9)$$

Similarly, the symmetric spin triplet (6) must pair to the antisymmetric spatial wavefunction (8), giving the overall antisymmetric triplet state $\Psi_{triplet}(\vec{r_1}, m_{s1}; \vec{r_2}, m_{s2})$,

$$\Psi_{triplet} = \frac{1}{\sqrt{2}}\left[\psi_{100}(\vec{r}_1)\psi_{nlm}(\vec{r}_2) - \psi_{100}(\vec{r}_2)\psi_{nlm}(\vec{r}_1)\right]$$
$$\times \frac{1}{\sqrt{3}}\left[\chi_{++} + \frac{1}{\sqrt{2}}(\chi_{+-} + \chi_{-+}) + \chi_{--}\right]. \quad (10)$$

The system can be described in terms of the quantum numbers for orbital angular momentum ($l$), magnetic quantum number ($m_l$), spin angular momentum ($s$), and spin quantum number ($m_s$). In addition to the before mentioned quantum numbers comes the principle quantum number ($n$), which is associated with the energy of the electron orbital. Starting with the principle quantum number, which can take any nonzero, positive integer value ($n = 1, 2, 3,...$), we are able to define the allowed values of the angular momentum and magnetic quantum number as follows (Liboff 1998):

$$l = 0, 1, 2,..., (n-1) \tag{11}$$

$$m_l = -l, -l + 1, ..., 0, 1, 2,..., +l. \tag{12}$$

In addition to the limitations placed on the possible states of angular momentum and the corresponding magnetic quantum numbers, there are quantum rules for the combining angular momenta. The reasons for adding angular momenta at the quantum level could entail the need to consider multiple particles within a system, or even the combination of different forms of angular momenta. Both of these scenarios will affect our quantum mechanical discussion of PDT. If we begin by defining a generic angular momentum term, $j$, two angular momenta ($j_1$ and $j_2$) can be added to reach the following permitted values:

$$j_{min} = |j_1 - j_2| \tag{13}$$

$$j_{max} = j_1 + j_2. \tag{14}$$

Based on these maximum and minimum values of total angular momenta,

$$j = |j_1 - j_2|, ..., j_1 + j_2 \tag{15}$$

is the range of acceptable total angular momenta values (Liboff 1998). When dealing with the addition of angular momenta, the range of

$$m_j = -(j_1 + j_2), ,..., 0, ..., j_1 + j_2 \tag{16}$$

follows from (12) and (15).

These allowed values for the quantum numbers are based on the solution for the spatial wavefunction of the hydrogen atom in spherical coordinates by separating radial and angular dependencies

$$\Phi(r, \theta, \phi) = R(r) Y_l^{m_l}(\theta, \phi), \tag{17}$$

where $R(r)$ represents the radial wavefunction and $Y_l^{m_l}(\theta, \phi)$ the spherical harmonics. Of key importance is the orthonormality of these special functions. Stating the wavefunction in terms of the given quantum numbers via subscripts, $\Phi_{nlm_l}$, taking the inner product of two such wavefunctions (or equivalently, integrating the product of the two wave functions over all space) returns

$$\langle \Phi_{n'l'm_l'} | \Phi_{nlm_l} \rangle = \delta_{nn'}\delta_{ll'}\delta_{m_l m_l'}, \tag{18}$$

where any given delta function takes the value of zero when the respective indices differ and unity when they are the same (Liboff 1998).

As an example illustrating the principle of conservation of energy, since the quantum number $n$ is tied to the energy of a state, the inner product of the final ($\Phi_{n'l'm_l'}$) and initial ($\Phi_{nlm_l}$) states will be zero if $n' \neq n$, meaning the system cannot transition spontaneously and unperturbed between the two states. The result will be unity if $n' = n$, such that the transition between the two states does not violate the conservation of energy. The only way to change the energy of the system between the initial and final states is to operate on them by doing work on the system, or by letting the system itself do work. Since there is no operator acting on the energy of the states in (18), the energy of the system must remain the same between the final and initial states.

Similarly, the conservation of angular momentum is thus upheld in reference to the angular momentum quantum number $l$ between the two states. A transition from $\Phi_{nlm_l}$ directly to $\Phi_{n'l'm_l'}$ is forbidden unless $l' = l$. This is of fundamental importance for the following discussion, as we shall see that the angular momentum of the PS goes from $l = 0$ to $l = 1$ during activation in PDT. This transition is however perfectly acceptable as the PS is being acted on by the incident light—by absorbing a photon (which carries an angular momentum of $l = 1$), the PS gains angular momentum in addition to energy. Later, this angular momentum will be transferred to molecular oxygen along with energy in order to elicit a phototoxic effect. Ultimately, when operating on one quantum state in order to cause it to transition to another quantum state, the operator acting on the system will invoke a set of selection rules as to which quantum transitions are allowed versus forbidden.

One further note should be made on the notation employed. The spin angular momentum ($s$) and spin quantum number ($m_s$) have been left out of the above conversation. However, as the name suggests, spin angular momentum is another form of angular momentum, or at least behaves quantum mechanically in the exact fashion as does angular momentum. The addition of spin angular momenta therefore abides the general rules for addition of angular momenta (15). The spin angular momentum of an electron is $s = \frac{1}{2}$, such that the associated spin quantum numbers are $m_s = \pm\frac{1}{2}$. Since both the PS and molecular oxygen of interest can each be considered two electron systems, their respective spin angular momenta can take values of $s = 0, 1$ via the rules for addition of angular momenta. Therefore, the spin quantum numbers for each of these individual molecules can take the values $m_s = 0, \pm 1$. The photon carries no spin angular momentum ($s = 0, m_s = 0$).

To begin our formal discussion of the quantum mechanical processes involved in PDT, we can use the addition of angular momenta in order to determine the state of each of the molecules using the condensed the notation

$$|\Psi\rangle_{molecule} = |l, s; m_l, m_s\rangle. \tag{19}$$

In this notation, the total state of the system is the product of the two molecular states

$$\begin{aligned}|\Psi\rangle_{system} &= |\Psi\rangle_{PS} \otimes |\Psi\rangle_O \\ &= |l, s; m_l, m_s\rangle_{PS} \otimes |l, s; m_l, m_s\rangle_O\end{aligned}, \tag{20}$$

where again $PS$ and $O$ refer to the photosensitizer and molecular oxygen, respectively.

Initially, both the PS and oxygen reside in their ground states—the PS in a spin singlet and the molecular oxygen a spin triplet (Fig. 2a)—such that

$$|\Psi\rangle_i = |l = 0, s = 0; m_l = 0, m_s = 0\rangle_{PS} \otimes |l = 0, s = 1; m_l = 0, m_s = 0\rangle_O. \tag{21}$$

Again using the addition of angular momentum, this time between the two molecules, the overall initial state given in terms of the same quantum numbers becomes

$$|\Psi\rangle_i = |l = 0, s = 1; m_l = 0, m_s = 0, \pm 1\rangle. \tag{22}$$

When the PS absorbs light of the appropriate wavelength, it transitions to an excited singlet state (Fig. 2b). Since the photon carries a quantum angular momentum of $l = 1$, this transition corresponds to an increase in orbital angular momentum of $\Delta l = +1$ within the PS. The state of molecular oxygen remains unchanged during this process. Upon absorption, the system transitions to the state

$$|\Psi\rangle_{abs} = |l = 1, s = 0; m_l = 0, \pm 1, m_s = 0\rangle_{PS} \otimes |l = 0, s = 1; m_l = 0, m_s = 0\rangle_O, \tag{23}$$

where again the addition of angular momentum between molecules gives the overall state

$$|\Psi\rangle_{abs} = |l = 1, s = 1; m_l = 0, \pm 1, m_s = 0, \pm 1\rangle. \tag{24}$$

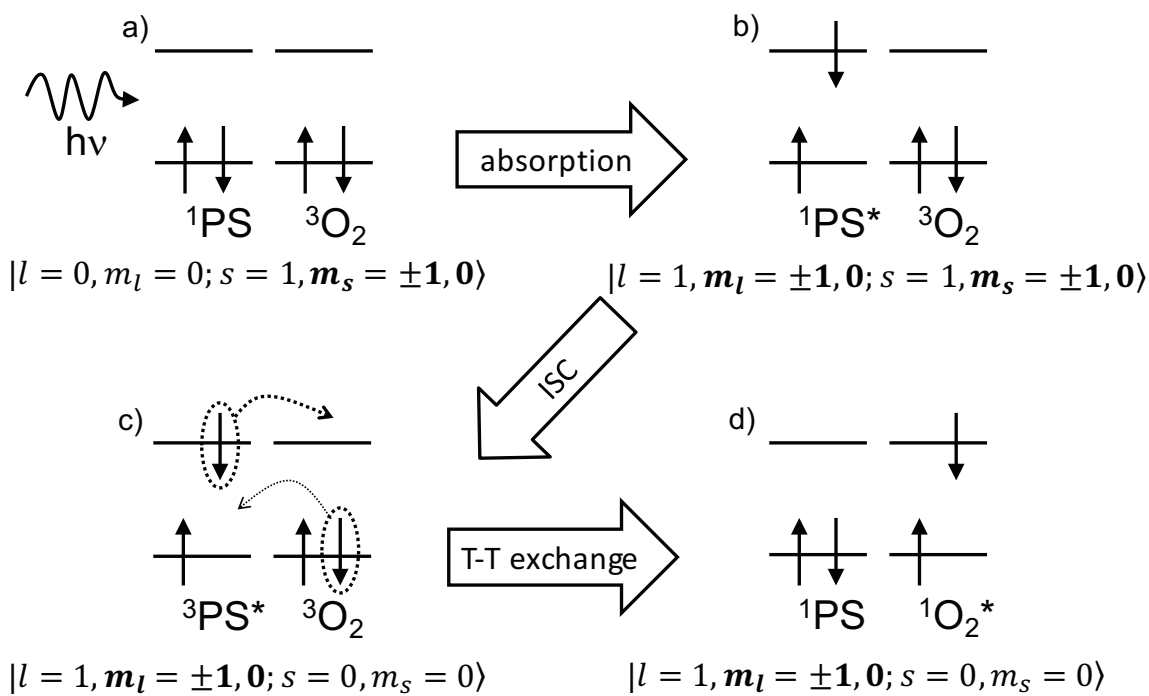

Figure 2. Energy level diagrams of the PDT process leading to the creation of singlet oxygen, depicted in a HOMO-LUMO representation. a) The initial states of the PS and molecular oxygen. b) The PS transitions to an excited spin singlet state via absorption. c) The PS transitions to an excited spin triplet state via Intersystem Crossing. d) Triplet-Triplet electron exchange between the PS and molecular oxygen leads to the final state of the system where the excited spin singlet state of oxygen is ready to impose oxidative damage in surrounding organisms.

Once in the excited state, the PS can either transition back to its ground state via fluorescence, or undergo a nonradiative transition to a spin triplet state. The later process is desirable for the PDT process, allowing the PS in its excited triplet state to interact with molecular oxygen. The nonradiative process by which the PS moves from an excited spin singlet to an excited spin triplet state is known as Intersystem Crossing, whereby the spin of the excited electron is no longer paired to that of the electron in the ground state (Fig. 2c) (Bonnett 2000). Due to the conservation of spin angular momentum, the transition from a singlet to a triplet state is a quantum mechanically forbidden transition. However, Intersystem Crossing is made possible by spin-orbit coupling, where the orbital and spin angular momenta are combined to give possible total angular momenta given in (15). This nonradiative transition relies upon the overlap of the vibrational states of the initial and final states of the electron (Bonnett 2000; Sakurai 1994; Liboff 1998; Beljonne et al. 2001} Again, molecular oxygen remains in its ground state during this process. Via Intersystem Crossing, the system transitions to the state

$$|\Psi\rangle_{ISC} = |l=1, s=1; m_l = 0, \pm 1, m_s = 0\rangle_{PS} \otimes |l=0, s=1; m_l = 0, m_s = 0\rangle_O. \qquad (25)$$

The addition of angular momentum between molecules gives the possible states

$$\begin{aligned}|\Psi\rangle_{ISC} =& |l=1, s=0; m_l = 0, \pm 1, m_s = 0\rangle \\ &+ |l=1, s=1; m_l = 0, \pm 1, m_s = 0, \pm 1\rangle \\ &+ |l=1, s=2; m_l = 0, \pm 1, m_s = 0, \pm 1, \pm 2\rangle,\end{aligned} \qquad (26)$$

where the states $s = 0, 1, 2$ are allowed along with their corresponding $-s \leq m_s \leq s$ values. Although the excited spin triplet state of the PS may phosphoresce back to its ground state, this state has a long-lived life time such that interaction with molecular oxygen becomes more likely (Hatz, Poulsen and Ogilby 2008).

The PS in its excited triplet state interacts with the molecular oxygen in its ground state (spin triplet) via a Triplet-Triplet Exchange of electrons (Fig. 2d). In this process, the excited electron of the PS transitions to the molecular oxygen and the electron with matching spin in the ground state of molecular oxygen transitions to the ground state of the PS. Along with this swapping of electrons comes an exchange of energy, such that the PS returns to its ground (spin singlet) state and the molecular oxygen transitions to an excited (spin singlet) state (Fig. 2d (Bonnett 2000;

Dexter 1953). The Triplet-Triplet Exchange is also referred to as a Dexter Exchange, based upon the seminal work "A Theory of Sensitized Luminescence in Solids" written by D.L. Dexter, which thoroughly explains this process. While the focus of this section is to simply give a general description of the quantum states of the PS and molecular oxygen during the stages of PDT, the reader is referred to Dexter's work for a more rigorous and thorough description of the exchange (Dexter 1953).

Continuing with the same quantum numbers, the corresponding wave function for the system becomes

$$|\Psi\rangle_{TT} = |l=0, s=0; m_l=0, m_s=0\rangle_{PS} \otimes |l=1, s=0; m_l=0, \pm 1, m_s=0\rangle_O. \qquad (27)$$

The addition of angular momentum between molecules gives the state

$$|\Psi\rangle_{TT} = |l=1, s=0; m_l=0, \pm 1, m_s=0\rangle. \qquad (28)$$

Given that the final state of this system must remain unchanged from that of (26) during this process, we can conclude that after the PS underwent Intersystem Crossing the system must have been in the first of those states listed in (26),

$$|\Psi\rangle_{ISC} = |l=1, s=0; m_l=0, \pm 1, m_s=0\rangle. \qquad (29)$$

From this conclusion, it follows that after the PS undergoes Intersystem Crossing, the system must be described by the individual molecular states

$$|\Psi\rangle_{ISC} = |l=1, s=1; m_l=0, \pm 1, m_s=0\rangle_{PS} \otimes |l=0, s=1; m_l=0, m_s=0\rangle_O. \qquad (30)$$

Pay particular attention to tracking the transfer and conservation of angular momentum throughout the processes described. Both molecules are in their ground states initially. Upon excitation of the PS via absorption of a photon, the angular momentum of the system increases.

While the angular momentum of the PS does not change during Intersystem Crossing, it does change in the final step as the angular momentum of the PS is transferred to that of the molecular oxygen. To better demonstrate the point, the final step of Figure 2—the triplet-triplet exchange between the PS and oxygen—is repeated again in Figure 3 along with the associated molecular orbitals of oxygen and protoporphyrin IX (PpIX), a typical photosensitizer employed clinically in PDT of cancers. The increase in angular momentum of molecular oxygen via the triplet-triplet exchange is visually apparent.

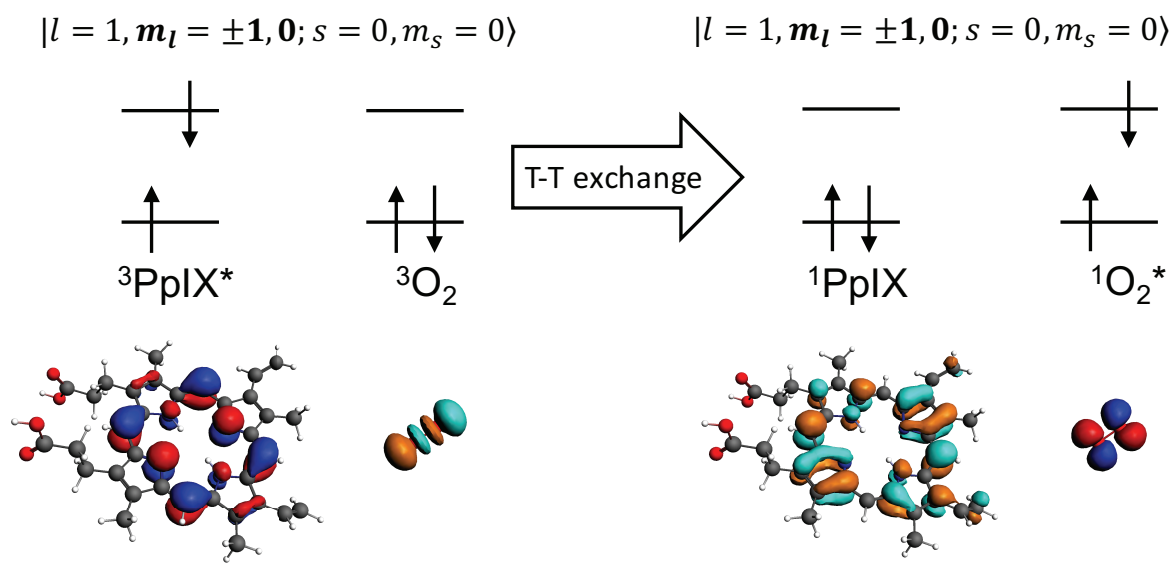

Figure 3. The HOMO-LUMO representations employed in the final step of Figure 2 are represented here again, with the corresponding molecular orbitals of $O_2$ and a common PS, protoporphyrin-IX (PpIX). Molecular orbitals were generated via the Amsterdam Density Functional program (te Velde et al. 2001).

SUMMARY

A summary of the states of the PS—$O_2$ system based upon the transitions and physical processes described would look as follows:

1. The PS and molecular oxygen begin in their ground states, the PS in a spin singlet and the molecular oxygen a spin triplet,

$$|\Psi\rangle_i = |l=0, s=0; m_l=0, m_s=0\rangle_{PS} \otimes |l=0, s=1; m_l=0, m_s=0\rangle_O. \tag{31}$$

2. Upon absorption of a photon, the PS is raised to an excited spin singlet state, while the molecular oxygen goes unaffected,

$$|\Psi\rangle_{abs} = |l=1, s=0; m_l=0,\pm 1, m_s=0\rangle_{PS} \otimes |l=0, s=1; m_l=0, m_s=0\rangle_O. \tag{32}$$

3. The PS undergoes a nonradiative transition from the excited spin singlet to an excited spin triplet via Intersystem Crossing, while the state of the molecular oxygen again remains unchanged in its spin triplet ground state,

$$|\Psi\rangle_{ISC} = |l=1, s=1; m_l=0,\pm 1, m_s=0\rangle_{PS} \otimes |l=0, s=1; m_l=0, m_s=0\rangle_O. \tag{33}$$

4. Finally, the molecular oxygen is raised from its ground spin triplet state to an excited spin singlet state as the PS simultaneously relaxes back from its excited spin triplet state to its spin singlet ground state,

$$|\Psi\rangle_{TT} = |l=0, s=0; m_l=0, m_s=0\rangle_{PS} \otimes |l=1, s=0; m_l=0,\pm 1, m_s=0\rangle_O. \tag{34}$$

Again, a summary of these processes and states is also depicted in Figure 2, where the overall wavefunction of the system at each step is listed with the corresponding energy diagram.

Simply put, energy from the excitation light is absorbed by the PS. Following some internal transitions, the PS is then able to transfer the added energy to the molecular oxygen via Triplet-Triplet Exchange. The final state of the PS—$O_2$ system leaves the molecular oxygen in an excited state, ready to unleash oxidative stress on its immediate surroundings, ultimately causing potential lethal photodamage as a result of biologic interactions that lead to activation of cellular death pathways (Finkel and Holbrook 2000; Martindale and Holbrook 2002; Pisoschi and Pop 2015; Apel and Hirt 2004).

ACKNOWLEDGMENTS

This publication was supported by an Institutional Development Award (IDeA) from the National Institute of General Medical Sciences of the National Institutes of Health under grant number P20 GM103418. The author would like to thank Henri J.F. Jansen for his advice while working through the details of this paper.

FIGURES

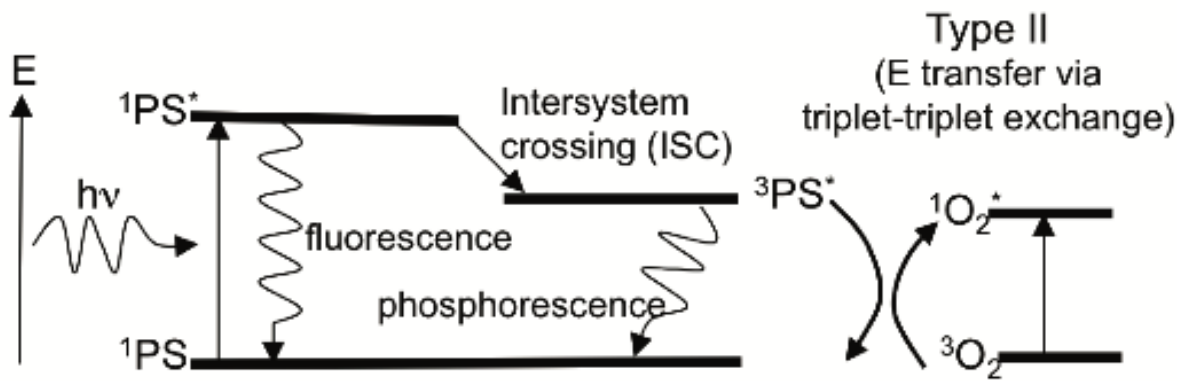

Figure 1. The process leading to the preferred Type II path to photodamage starts when the PS is excited by incident light of energy $h\nu$. The PS then relaxes via ISC to an excited triplet state, whereby it can transfer energy to molecular oxygen via a triplet-triplet electron transfer.

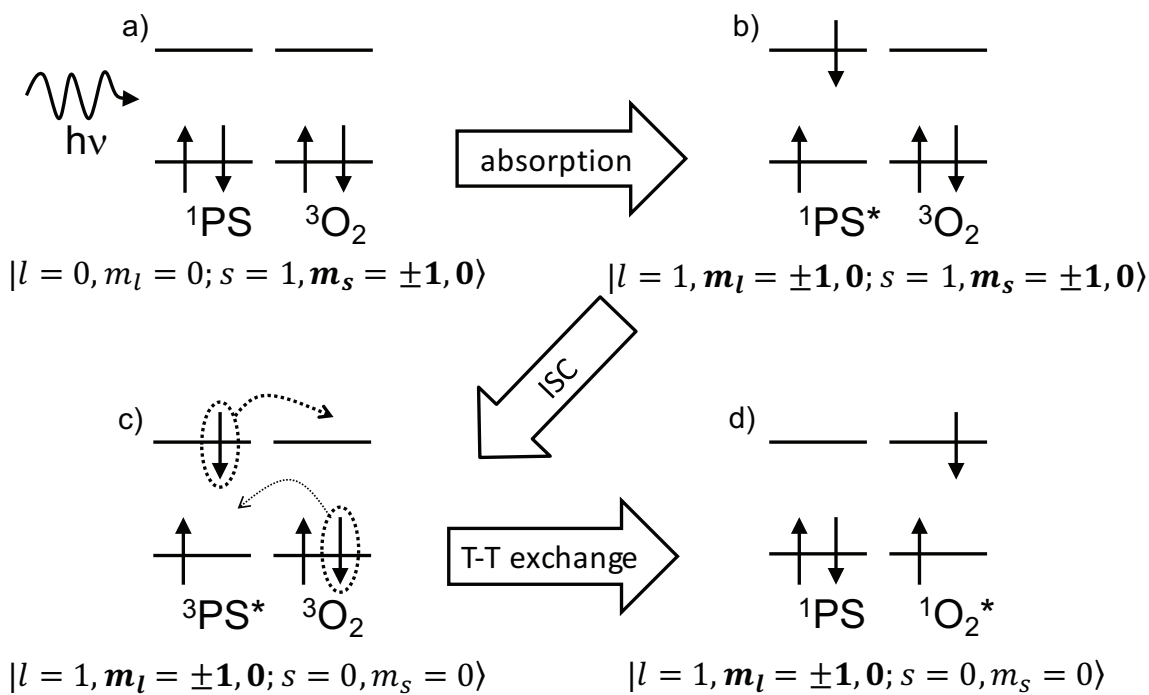

Figure 2. Energy level diagrams of the PDT process leading to the creation of singlet oxygen, depicted in a HOMO-LUMO representation. a) The initial states of the PS and molecular oxygen. b) The PS transitions to an excited spin singlet state via absorption. c) The PS transitions to an excited spin triplet state via Intersystem Crossing. d) Triplet-Triplet electron exchange between the PS and molecular oxygen leads to the final state of the system where the excited spin singlet state of oxygen is ready to impose oxidative damage in surrounding organisms.

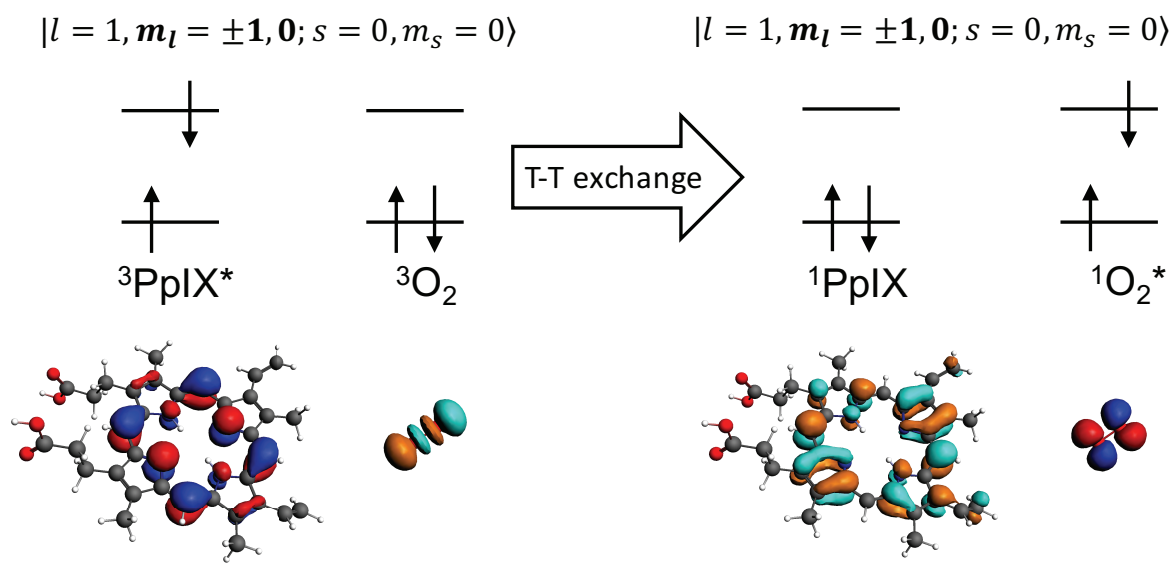

Figure 3. The HOMO-LUMO representations employed in the final step of Figure 2 are represented here again, with the corresponding molecular orbitals of $O_2$ and a common PS, protoporphyrin-IX (PpIX). Molecular orbitals were generated via the Amsterdam Density Functional program (te Velde et al. 2001).